\documentclass[showpacs,twocolumn,preprintnumbers,amsmath,amssymb,article,epsf,superscriptaddress]{revtex4-1}
\usepackage{graphicx}
\usepackage{nicefrac}
\usepackage[squaren]{SIunits}
\usepackage{color}

\begin{document}
\title{Multiphoton Effects Enhanced Due to Ultrafast Photon-Number Fluctuations}

\author{Kirill Yu.~Spasibko}
\affiliation{Max Planck Institute for the Science of Light,
Staudtstrasse 2, 91058 Erlangen, Germany}
\affiliation{University of Erlangen-N\"urnberg, Staudtstrasse 7/B2, 91058 Erlangen, Germany}
\author{Denis A.~Kopylov}
\affiliation{Department of Physics, M.V.Lomonosov Moscow State University, \\ Leninskie Gory, 119991 Moscow,
Russia}
\author{Victor L.~Krutyanskiy}
\affiliation{Department of Physics, M.V.Lomonosov Moscow State University, \\ Leninskie Gory, 119991 Moscow,
Russia}
\affiliation{Institute for Quantum Optics and Quantum Information \"OAW, Technikerstr. 21a, 6020 Innsbruck, Austria}
\author{Tatiana V.~Murzina}
\affiliation{Department of Physics, M.V.Lomonosov Moscow State University, \\ Leninskie Gory, 119991 Moscow,
Russia}
\author{Gerd Leuchs}
\affiliation{Max Planck Institute for the Science of Light, Staudtstrasse 2, 91058
Erlangen, Germany}
\affiliation{University of Erlangen-N\"urnberg, Staudtstrasse 7/B2, 91058 Erlangen, Germany}
\author{Maria~V.~Chekhova}
\affiliation{Max Planck Institute for the Science of Light, Staudtstrasse 2, 91058
Erlangen, Germany}
\affiliation{University of Erlangen-N\"urnberg, Staudtstrasse 7/B2, 91058 Erlangen, Germany}
\affiliation{Department of Physics, M.V.Lomonosov Moscow State University, \\ Leninskie Gory,
119991 Moscow, Russia}

\vspace{-10mm}

\vspace{5mm}
\maketitle \narrowtext

\textbf{Multi-photon processes are the essence of nonlinear optics. Optical harmonics generation \cite{Franken1961} and multi-photon absorption \cite{Denk1990}, ionization \cite{Delone2000}, polymerization \cite{Gattass2008} or spectroscopy \cite{Shen1984} are widely used in practical applications. Generally, the rate of an $n$-photon effect scales as the $n$-th order autocorrelation function of the incident light \cite{Mollow1968, Agarwal1970,Lecompte1975,Delone1980,Qu1992,Qu1995,Popov2002,Jechow2013}, which is high for light with strong photon-number fluctuations. Therefore `noisy' light sources are much more efficient for multi-photon effects than coherent sources with the same mean power, pulse duration and repetition rate. Here we generate optical harmonics of order 2-4 from bright squeezed vacuum (BSV)~\cite{Jedrkiewicz2004,Iskhakov2009,Agafonov2010,Corzo2011}, a state of light consisting of only quantum noise with no coherent component. We observe up to two orders of magnitude enhancement in the generation of optical harmonics due to ultrafast photon-number fluctuations. This feature is especially important for the nonlinear optics of fragile structures where the use of a `noisy' pump can considerably increase the effect without overcoming the damage threshold.}

In any multi-photon process, one example being the generation of optical harmonics, a certain number of photons of the initial radiation get annihilated to produce the effect. For this reason, the output signal of an $n$-photon process scales as the normally ordered $n$-th order moment of the incident photon number, $\langle:N^n:\rangle\equiv\langle(a^\dagger)^na^n\rangle$~\cite{Mollow1968,Agarwal1970}, similar to the rate of $n$-photon coincidences~\cite{Glauber1963} in a quantum optics experiment. It follows that the rate $R^{(n)}$ of an $n$-photon effect scales as the $n$-th order normalized autocorrelation function of the incident light, $g^{(n)}\equiv\langle:N^n:\rangle/\langle N\rangle^n$,
\begin{equation}
R^{(n)}\sim  g^{(n)}F^n,
\label{eq:multisignal}
\end{equation}
where $F\sim\langle N\rangle$ is the input photon flux.

According to Eq.~(\ref{eq:multisignal}), one can define the \textit{statistical efficiency} of an $n$-photon effect~\cite{Qu1992,Lecompte1975,Delone1980} as
\begin{equation}
\xi^{(n)}\equiv \frac{R^{(n)}}{F^n}.
\label{eq:efficiency}
\end{equation}
In this form the efficiency follows the changes in the statistics of the input radiation and does not depend on its photon flux. Clearly, it should scale with the $n$-th order normalized intensity correlation function.

Enhancement due to intensity fluctuations has been demonstrated for the second-harmonic generation~\cite{Qu1992,Qu1995} and recently, also for two-photon absorption~\cite{Jechow2013} from thermal light, for which $g^{(n)}=n!$. Instead of thermal light, one can use multimode laser light~\cite{Lecompte1975,Delone1980} as the latter has intensity fluctuations due to the contributions of different temporal modes. In fact, pulsed light is more efficient for $n$-photon processes than continuous-wave light with the same mean intensity for the same reason: it has strong intensity modulation.

Nevertheless, in practice one always observes multiphoton effects with coherent light, for which $g^{(n)}=1$. The reason is that thermal light with high intensity is hard to produce; thermal radiation used in Refs.~\cite{Qu1992,Qu1995,Jechow2013} was obtained from a below-threshold laser, which is not very bright. Another standard source of thermal light, rotating ground glass disc, not only suffers from low brightness, but also has very slow intensity fluctuations. As an efficient source, one would desire a bright pulsed one with each pulse having ultrafast intensity fluctuations.

\begin{figure*}[tb]
\begin{center}
\includegraphics[width=0.8\textwidth]{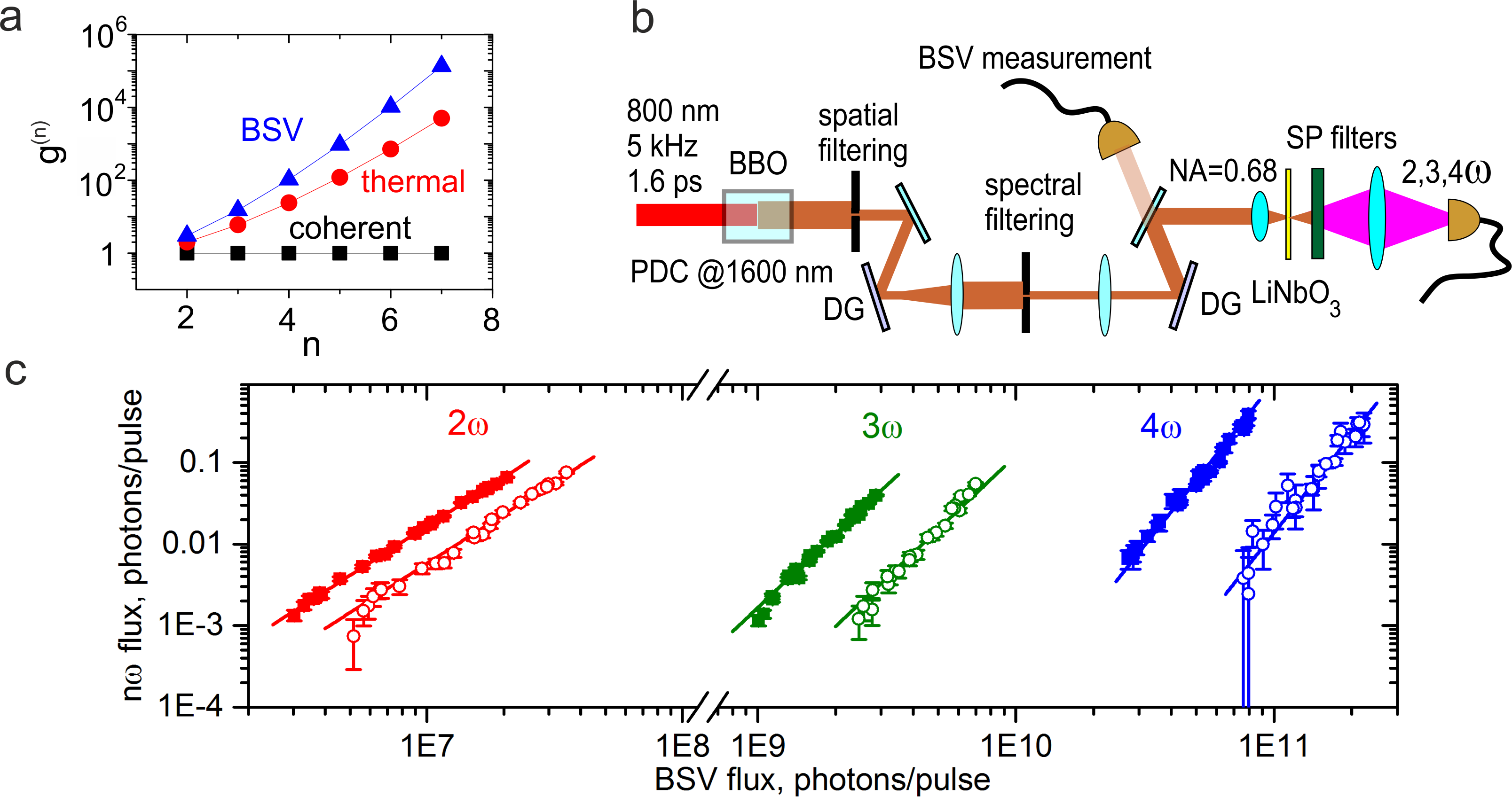}
\caption{(a) The normalized correlation function determining the enhancement of an $n$-photon effect as a function of $n$ in the case of coherent (black), thermal (red) and BSV (blue) light used as a pump. (b) The experimental setup. Bright squeezed vacuum is obtained via high-gain PDC in a BBO crystal and filtered spatially with a slit and spectrally with a 4f-system containing two diffraction gratings (DG), two lenses and a slit. After the filtering, a small part of BSV is tapped off for monitoring the mean photon number per pulse and the statistics, and the rest is tightly focused on the surface of a lithium niobate crystal (LiNbO$_3$). The second, third, and fourth optical harmonics are generated without phase matching, filtered from the BSV radiation with short-pass (SP) filters and registered by a detector. (c) Measured dependence of the SH (red), TH (green) and FH (blue) photon flux on the pump flux with the pump being BSV (filled squares) and pseudo-coherent light obtained from BSV through post-selection (empty circles). In each case the flux is corrected for the detection and transmission losses. Theoretical fits with the quadratic, cubic, and quartic functions (lines) show the correct power dependence for each harmonic.}
\label{fig:efficiency}
\end{center}
\end{figure*}

Exceptionally interesting in this connection is bright squeezed vacuum (BSV) produced via high-gain parametric down-conversion (PDC). It can be extremely bright --- up to hundreds of mW mean power \cite{Perez2015,Spasibko2016} --- and has highly fluctuating photon number. In fact, this state has no coherent component and only consists of quantum noise. Taken separately, signal and idler BSV beams have thermal statistics; but whenever they are indistinguishable, the statistics shows `superbunched' behavior with $g^{(n)}=(2n-1)!!$ \cite{Janszky1987,Boitier2011,Iskhakov2012}. This makes BSV extremely efficient for multi-photon effects \cite{Popov2002}, as one can see from Fig.~\ref{fig:efficiency}a showing $g^{(n)}(n)$ for the cases of coherent (black), thermal (red) and BSV (blue) light.

BSV intensity fluctuations are not only strong; they can be extremely fast as well. Due to the broad spectrum, whose bandwidth can reach  100-150 THz \cite{Strekalov2005,Tiihonen2006,Katamadze2015}, BSV generated through PDC has correlation times as short as a few femtoseconds. For pulsed broadband BSV the intensity will fluctuate not only from pulse to pulse, but also within a single pulse. Although pulse-to-pulse fluctuations could be, in principle, mimicked by modulating the beam, fluctuations within the pulse can not.

The latter makes BSV an extremely valuable source for nonlinear effects. Any nonlinear effect with ultrashort response time will `feel' the fast photon-number fluctuations of BSV and its efficiency will be therefore dramatically enhanced. In our experiment, we demonstrate this enhancement in the generation of the second (SH), third (TH) and fourth (FH) optical harmonics.

The experimental setup is shown in Fig.~\ref{fig:efficiency}b. For the generation of optical harmonics we use broadband BSV radiation around the frequency degeneracy (1600 nm) \cite{Spasibko2016} produced in a beta barium borate (BBO) crystal (see Methods). After spatial and spectral filtering, BSV is tightly focused on the surface of a lithium niobate (LiNbO$_3$) crystal, where SH, TH, and FH are generated without phase-matching. The radiation of optical harmonics is separated from BSV first by short-pass filters and then by different bandpass filters. The photon number in each pulse of the harmonics radiation is measured with an avalanche photodiode or with a charge-integrating detector based on a Si PIN photodiode~\cite{Iskhakov2009}.

For monitoring and controlling the statistics of BSV radiation, we tap off its small part and measure the numbers of photons in each pulse with an infrared charge-integrating detector under narrowband filtering (see Methods).

The values of normalized correlation functions measured for BSV indeed coincide, within the measurement accuracy, with the theoretical predictions (see Methods). However, these values can be achieved not at the highest possible power of BSV. The reason is that inside the BBO crystal producing PDC, incoherent second-harmonic generation takes place if the PDC power is high. This process is more likely to happen for strongest pulses; therefore~\cite{Krasinski1976,Leuchs1986} it reduces photon-number fluctuations and leads to a decrease in $g^{(n)}$. We use this fact to control the statistics of BSV: by changing the phase matching and, correspondingly, by making BSV radiation weaker or stronger, we make $g^{(n)}$ higher or lower, respectively (see Methods for more details).

For each optical harmonic, we measure the dependence of the output photon flux on the BSV flux, spectrally filtered to a bandwidth of $3.3$ nm. Panel c in Fig.~\ref{fig:efficiency} shows these dependences by red, green, and blue filled squares for the second, third, and fourth harmonics, respectively.

\begin{table}
\begin{tabular}{|c|c|c|c|}
\hline
n&$A^{(n)}_{BSV}/A^{(n)}_{ps}$&$g^{(n)}_{BSV}/g^{(n)}_{ps}$&$\eta^{(n)}_{\max}$\\
\hline
2&$2.86\pm0.08$&$2.94\pm0.06$&$(3.2\pm0.3)\times10^{-7}\%$\\
\hline
3&$13.6\pm0.8$&$14.5\pm0.6$&$(1.38\pm0.12)\times10^{-9}\%$\\
\hline
4&$71\pm6$&$63\pm2$&$(4.9\pm0.7)\times10^{-10}\%$\\
\hline
\end{tabular}

\caption{Characteristics of the obtained harmonic generation from BSV: statistical enhancement with respect to the pseudo-coherent source ($A^{(n)}_{BSV}/A^{(n)}_{ps}$) (left), the ratios of the corresponding correlation functions $g^{(n)}_{BSV}/g^{(n)}_{ps}$ (center) and  maximal harmonic generation efficiency $\eta^{(n)}_{\max}\equiv F_{n\omega}/F$ from Fig.~\ref{fig:efficiency}c (right).}
\end{table}

The statistical efficiencies $\xi^{(n)}$, according to Eq.~\ref{eq:efficiency}, can be obtained from the fits with the quadratic, cubic, and quartic dependencies, $f_n(x)=A^{(n)}x^n$, $n=2,3,4$, shown by solid lines. This has to be compared with the `standard' situation where the harmonics are generated from coherent light. In the absence of a coherent source with exactly the same wavelength, pulse duration, and repetition rate as the BSV radiation, we create a pseudo-coherent source. To this end, we reduce the photon-number fluctuations of BSV using post-selection (see Methods). By imposing a condition on the photon number in the tapped off beam to be within certain boundaries, we also reduce the photon-number fluctuations in the transmitted beam~\cite{Iskhakov2016}, so that the intensity correlation functions for the post-selected pulses reach almost unity. 
 The power dependences for different harmonics generation from this pseudo-coherent source are plotted in Fig.~\ref{fig:efficiency}c by empty circles, with the same colours as in the case of BSV, and also fitted by power dependences (lines).
One can see that the statistical efficiency is indeed considerably higher for BSV than for pseudo-coherent light. The enhancement factors are shown in Table I (second column), together with
the ratios of the correlation functions for BSV and the pseudo-coherent light source (third column). As expected, there is agreement between the two.

\begin{figure}[h]
\begin{center}
\includegraphics[width=0.35\textwidth]{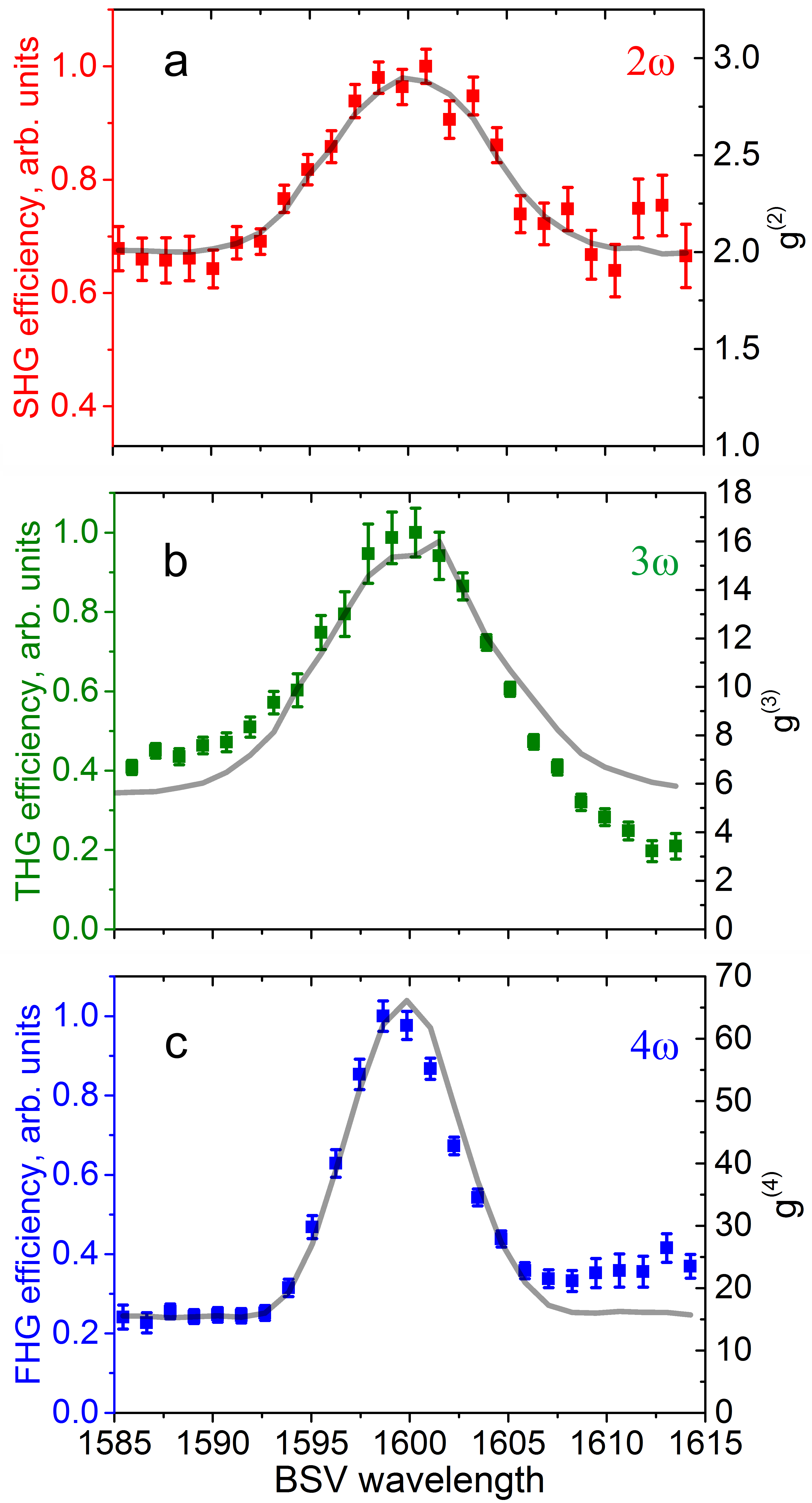}
\caption{SHG (a), THG (b) and FHG (c) statistical efficiencies $\xi^{(n)}$ (points) as well as the corresponding correlation functions (lines) measured versus the BSV wavelength.}
\label{fig:comp_thermal}
\end{center}
\end{figure}
To compare the efficiency of harmonics generation from degenerate BSV with the one from thermal light, we scan the BSV wavelength around the degeneracy point $1600$ nm and measure the rates of the SHG, around $800$ nm, of the THG, around $533$ nm, and of the FHG, around $400$ nm. 
The generation efficiencies, calculated according to Eq.~\ref{eq:efficiency}, are plotted in Fig. \ref{fig:comp_thermal}, together with the corresponding correlation functions of BSV. By scanning the wavelength we pass from BSV with thermal statistics (far from the degeneracy point) to BSV with the superbunched statistics (around the degeneracy point). The efficiency follows the correlation function, and we observe an enhancement by factors of $2$, $2.5$, and $4$ for SHG, THG, and FHG with BSV compared to thermal light, respectively.

\begin{figure*}[tb]
\begin{center}
\includegraphics[width=0.8\textwidth]{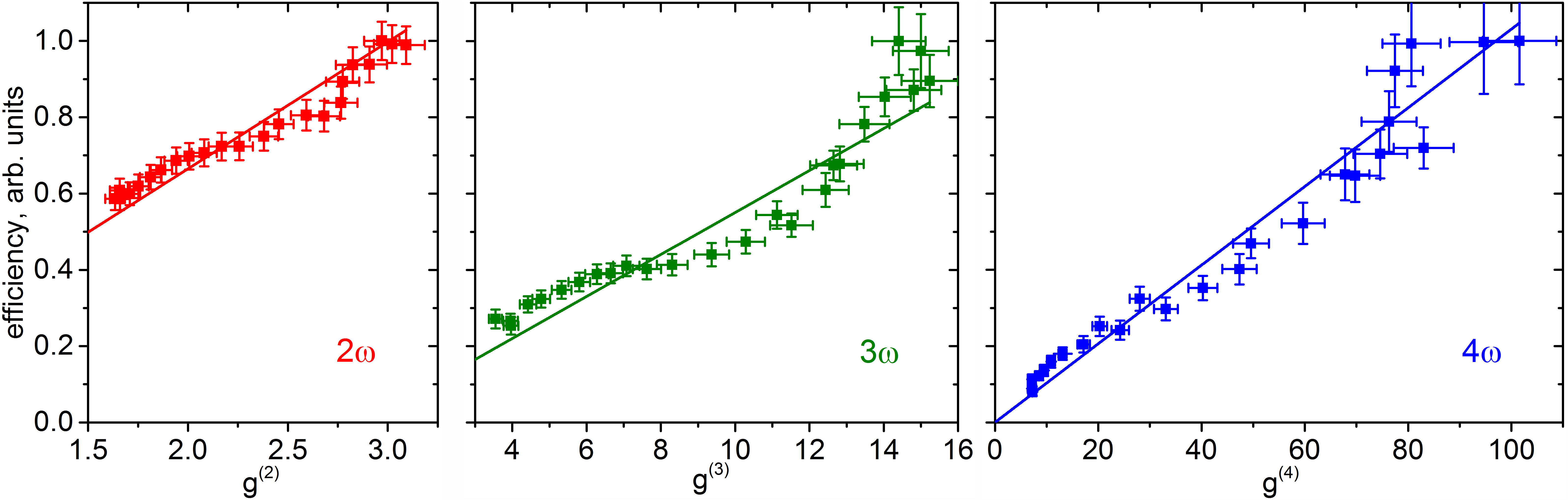}
\caption{Statistical efficiencies $\xi^{(n)}$ of SHG (left), THG (middle) and FHG (right) from broadband BSV show linear dependences on the corresponding correlation functions, measured with narrowband filtering. The correlation functions are varied by changing the orientation of the BBO crystal and hence the efficiency of BSV generation (see Methods). Low values of the correlation functions are obtained with the BSV generated efficiently, so that the second-harmonic generation in BBO reduces the photon-number fluctuations. In the opposite case, with inefficient BSV generation, its superbunched statistics is almost perfectly preserved, and the correlation functions are high.}
\label{fig:scaling}
\end{center}
\end{figure*}

The scaling of the harmonic generation efficiency with the normalized correlation functions is demonstrated by Fig.~\ref{fig:scaling}, which shows the efficiencies of SHG (left), THG (middle) and FHG (right) from BSV measured as functions of the second-, third-, and fourth-order correlation functions, respectively. To vary the values of the correlation functions we change the orientation of the BBO crystal, which changes the BSV generation efficiency and hence the statistics of the BSV through the second-harmonic generation in the same BBO crystal, see Methods.

In all measurements shown in Fig.~\ref{fig:scaling}, the optical harmonics are generated from broadband BSV, with the typical times of intensity fluctuations being $80$ fs. At the same time, $g^{(n)}$ were measured under $2.5$ nm filtering (see Methods).

Because the statistical efficiency $\xi^{(n)}$ of an $n$-th order harmonic generation scales as the $n$-th order normalized correlation function, the data presented in all three panels of Fig.~\ref{fig:scaling} generally follow a linear fit $\xi^{(n)}\sim g^{(n)}$ (lines). 

We see that the photon-number `noise' dramatically increases the efficiency of optical harmonic generation. It turns out that the generated harmonics are even `noisier' \cite{Ducuing1964}. Indeed, the second-order normalized correlation function $g^{(2)}_{n\omega}$ for the $n$-th harmonics is, by definition,
\begin{equation}
g^{(2)}_{n\omega}\equiv\frac{\langle:N_{n\omega}^2:\rangle}{\langle N_{n\omega}\rangle^2}.
\end{equation}
The normal ordering can be omitted due to the high mean photon number $\langle N_{n\omega}\rangle$. Then, from Eq.~(\ref{eq:multisignal}), we get~\cite{Teich1966,Qu1995}
\begin{equation}
g^{(2)}_{n\omega}=\frac{g^{(2n)}_{\omega}}{(g^{(n)}_{\omega})^2}.
\label{g_n_omega}
\end{equation}
Thus, the correlation functions for the harmonics can be estimated from the ones for the pump. Note that the obtained result is valid for any type of the pump statistics.

\begin{figure}[h]
\begin{center}
\includegraphics[height=0.25\textwidth]{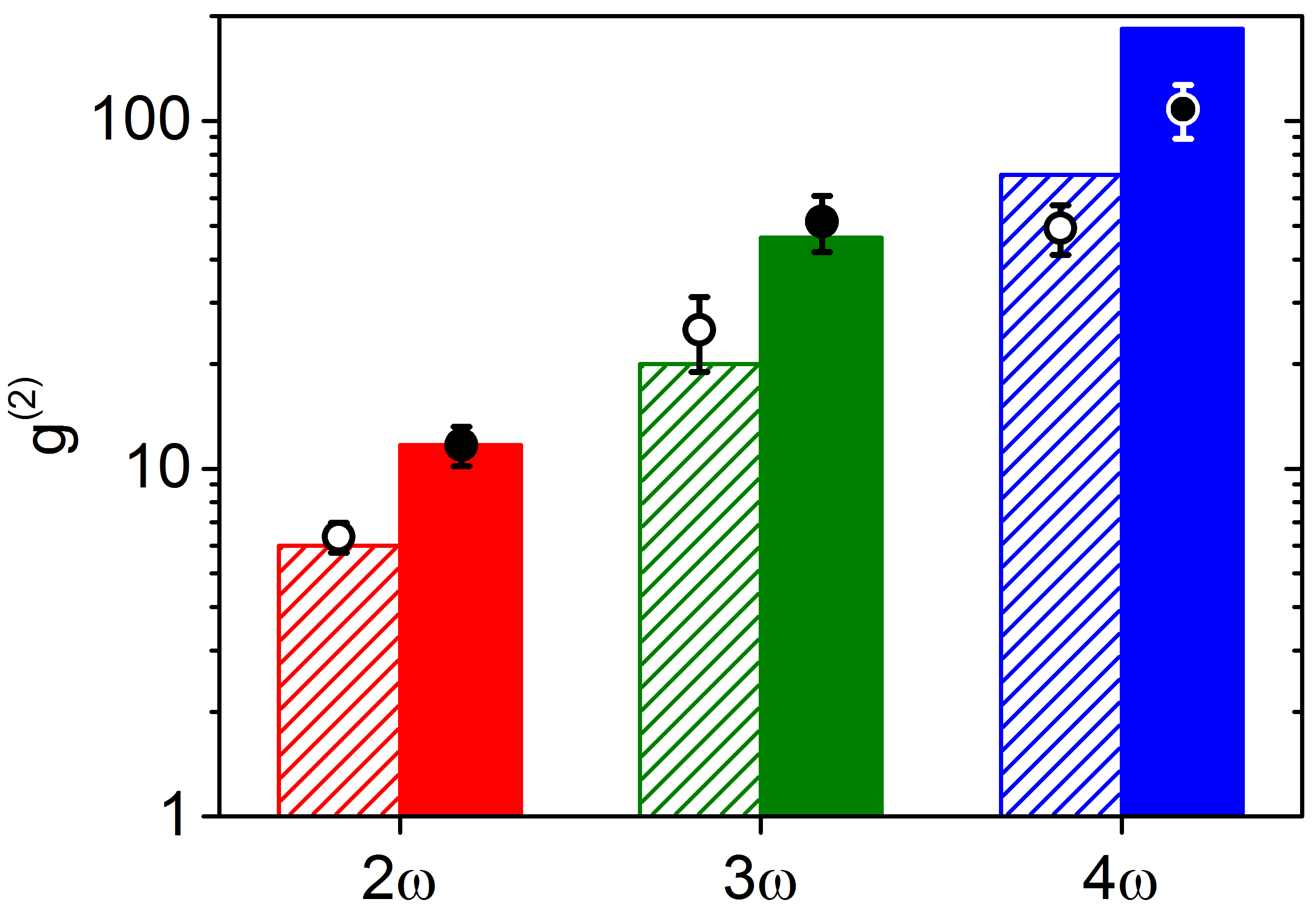}
\caption{Measured second-order autocorrelation functions $g^{(2)}_{n\omega}$ for the harmonics generated from BSV with superbunched (filled circles) and thermal (empty circles) statistics show extremely high fluctuations of the harmonic radiation, which increase with the number of the harmonic. The theoretical values are shown by filled and shaded colour bars, respectively.}
\label{fig:g2_harm}
\end{center}
\end{figure}

The bars in Fig.~\ref{fig:g2_harm} show the theoretically predicted values of the second-order normalized correlation function (bunching parameter) for all three optical harmonics generated from superbunched (colored) and thermal (shaded) BSV. The expected values are very high and range from $6$, for SHG from thermal light \cite{Qu1995,Allevi2015}, to $184$, for FHG from superbunched BSV.
The experimental results are shown by black circles on each bar. The correlation function was measured by registering coincidences in a Hanbury Brown-Twiss setup (see Methods). The data, in good agreement with the theory, show a huge increase of fluctuations in the generated optical harmonics.

These results, obtained for one particular case of optical harmonic generation, show that BSV is an extremely useful source for multi-photon effects in general. Among other applications, the use of BSV as a pump will be beneficial for the nonlinear optics of fragile structures. Indeed, in our experiment we generate harmonics without phase-matching from extremely low powers (from a few nW to tens of $\mu$W). According to Fig.~\ref{fig:efficiency}, to achieve the same rate of a four-photon effect one can use BSV with the mean power reduced about three times compared to the case of coherent pumping. This can help to dramatically increase the sensitivity while being considerably below the damage threshold.

An important feature of BSV is the time scale of its photon-number fluctuations. Here we have demonstrated a bandwidth of $9$ THz, corresponding to times of $80$ fs, but one can achieve times at least an order of magnitude shorter. Such behaviour cannot be mimicked by any external intensity modulation. Moreover, although here for simplicity we used a single spatial mode, the multimode spatial structure of PDC will not change its `noisy' behaviour and one can use the whole frequency and angular spectrum of BSV for exciting multiphoton effects.

Finally, the observation of extremely strong photon bunching in the radiation of optical harmonics from BSV is very interesting in connection with the intensely discussed topic of extreme events and `rogue waves'~\cite{Dudley2014}, observed for processes with `heavy-tailed' probability distributions. To the best of our knowledge, the observed $g^{(2)}=110\pm20$ considerably exceeds all values reported in the literature \cite{Meuret2015} and indicates a photon-number probability distribution with a very `heavy tail'. The perspective of observing multi-photon effects from such radiation is even more interesting; however, it would require a higher sensitivity than the one of our experiments.

\textbf{Acknowledgments.} We acknowledge the financial support of the joint DFG-RFBR project CH1591/2-1 - 16-52-12031 NNIOa and of the G-RISC program, projects P-2017a-19, P-2016a-5, P-2016a-6.

\textbf{Author contributions.} K.Yu.S., D.A.K. and V.L.K. carried out the experimental work and contributed to the analysis of the data; M.V.C., T.V.M. and G.L. initiated and supervised the project; all authors discussed
the results and contributed to the writing of the manuscript.

\textbf{Methods}

\textbf{BSV generation and controlling its statistics.} We produce broadband BSV radiation in a $10$ mm BBO crystal using as a pump the amplified radiation of a Ti-sapphire laser with the wavelength $800$ nm, the pulse duration $1.6$ ps, and the repetition rate $5$ kHz. The pump is focused into the crystal by a cylindrical lens with a focal length of $700$ mm, so that the beam remains unfocused in the plane of the optic axis and the pump walk-off~\cite{Perez2015} is not pronounced. The crystal is cut at 19.9$^{\circ}$ degrees to the optic axis, which provides type-I collinear phase-matching around the degenerate wavelength $1600$ nm~\cite{Spasibko2016}. After the crystal we cut off the pump by dichroic mirrors and two FEL1200 long-pass filters. A slit is used to filter BSV to a single spatial mode. The spectral filtering is made by a 4f-monochromator, which allows the wavelength to be scanned from $1550$ nm to $1650$ nm and the bandwidth around the degenerate wavelength $1600$ nm to be changed. Although the bandwidth of unfiltered BSV is $400$ nm~\cite{Spasibko2016}, the largest bandwidth allowed by filtering is $75$ nm.

We found out that BSV in the BBO crystal experiences further sum frequency generation, 1600 nm $\to$ 800 nm. This process acts on BSV as a nonlinear absorber. It reduces fluctuations by absorbing more photons from higher energy bursts~\cite{Krasinski1976,Leuchs1986}. The contribution of this process can be tuned by tilting the crystal closer or further from the exact phase-matching position. Using this method we change the statistics of BSV and can reduce its correlation functions to $g^{(2)}=1.55\pm0.05$, $g^{(3)}=3.5\pm0.2$, and $g^{(4)}=7.1\pm0.5$. The dependence of $g^{(2)}$ on the crystal orientation is shown in Fig.~5.

With the crystal oriented far from the exact phase matching, the strong photon-number fluctuations of BSV are maintained, and the measured correlation functions reach their theoretical values: $g^{(2)}_{BSV}=2.97\pm0.06$, $g^{(3)}_{BSV}=14.8\pm0.6$, and $g^{(4)}_{BSV}=108\pm9$ (Fig.~6). At the same time, the highest values of $g^{(n)}$ can be only achieved with the mean power of BSV not higher than $4$ mW before spatial and spectral filtering, which is not enough for FHG after all filtering losses. The fourth harmonic is therefore produced with a reduced value of $g^{(4)}=69\pm2$.

\textbf{Obtaining pseudo-coherent light by post-selection.} For controlling the statistics of BSV, we tap off about 0.6\% of its power on a beam sampler. For each pulse, we record the number of photons in the reflected BSV beam and in the harmonic radiation. From the obtained dataset we choose only pulses in which the number of photons in BSV lies within certain boundaries. Only for those pulses the data on the harmonics are further processed. This way we restrict the BSV fluctuations and mimic the generation of optical harmonics from coherent light. We choose this  post-selection `window' to be as small as possible, but large enough for the harmonic flux after post-selection to be measurable (Fig.~7). The minimal values of correlation functions obtained this way are $g^{(2)}=1.01\pm0.002$, $g^{(3)}=1.02\pm0.003$, $g^{(4)}=1.1\pm0.007$. 
After post-selection, the flux of each optical harmonic is linear in the corresponding correlation function; Fig.~8 shows THG as an example.

\textbf{Correlation functions for BSV} are measured under narrowband filtering, the bandwidth ranging from $4.5$ nm to $2.5$ nm. Otherwise, intensity fluctuations within each pulse lead to multimode detection and therefore to the reduction of the correlation functions. We measure the photon number in each BSV pulse using a charge-integrating detector with the same design as the one used in the visible range, but based on an InGaAs PIN photodiode Hamamatsu G12180-020A. The detector has a dynamic range from few thousands to several millions of photons per pulse, a quantum efficiency of 85\% at 1600 nm, and the electronic noise equivalent to $1600$ photons per pulse. It produces electronic pulses with the area $S=KN$, where $N$ is the input number of photons and $K=5.0\pm0.4\quad pV\times s$. The correlation functions are calculated from the data using the formula $g^{(n)}=M^{n-1}\sum_iS_i^n/\left[\sum_iS_i\right]^n$, where $S_i$ is the area of each electronic pulse and $M$ is the number of pulses. Here, one can use a single detector and ignore the normal ordering in the correlation function definition due to the large mean number of photons in BSV.

\textbf{Generation of optical harmonics.}
In the generation of all optical harmonics, we focus BSV on the surface of a $1$ mm thick lithium niobate crystal slab by a lens with the focal length $3.1$ mm and numerical aperture 0.68. Because the Rayleigh length ($7 \mu$m) is much smaller than the crystal length, harmonic generation in the bulk is suppressed due to the Gouy phase, and for more efficient harmonic generation the beam waist is placed on the crystal surface \cite{Rostovtseva1980}. The crystal optic axis is in the plane of the slab; this orientation does not provide phase matching for any harmonic generation. The polarization of the BSV as well as of all harmonics is along the optic axis, which means that the interaction involves the $z$ components of all nonlinear susceptibilities, which are the largest for LiNbO$_3$. For the filtering of optical harmonics, the following filters are used: three FGB25 and FBH800-40 for SH, two FGS900A and MF530-43 for TH, and two FGS900A and FBH400-40 for FH.

\textbf{Measurement of the second-order correlation function for the optical harmonics.}
For each harmonic generated from BSV after narrowband filtering, the correlation function $g^{(2)}$ is measured with a Hanbury Brown -- Twiss setup. The harmonic radiation is collimated by a lens with the focal length $3.1$ mm, split using a zero-order wave plate and a Glan prism, and fed in two gated single-photon Perkin\&Elmer SPCM-AQRH-16 counters. The photocounts are sent to a coincidence circuit and the correlation function is found as $g^{(2)}=M\sum_iN^{c}_i/\left[\sum_iN^{1}_i\sum_iN^{2}_i\right]$, where $N^{c,1,2}_i=0,1$ are the numbers of coincidences and counts in detectors 1 and 2, respectively, for different pulses and $M$ is the number of pulses. 

\begin{figure*}[h]
\begin{center}
\includegraphics[width=0.6\textwidth]{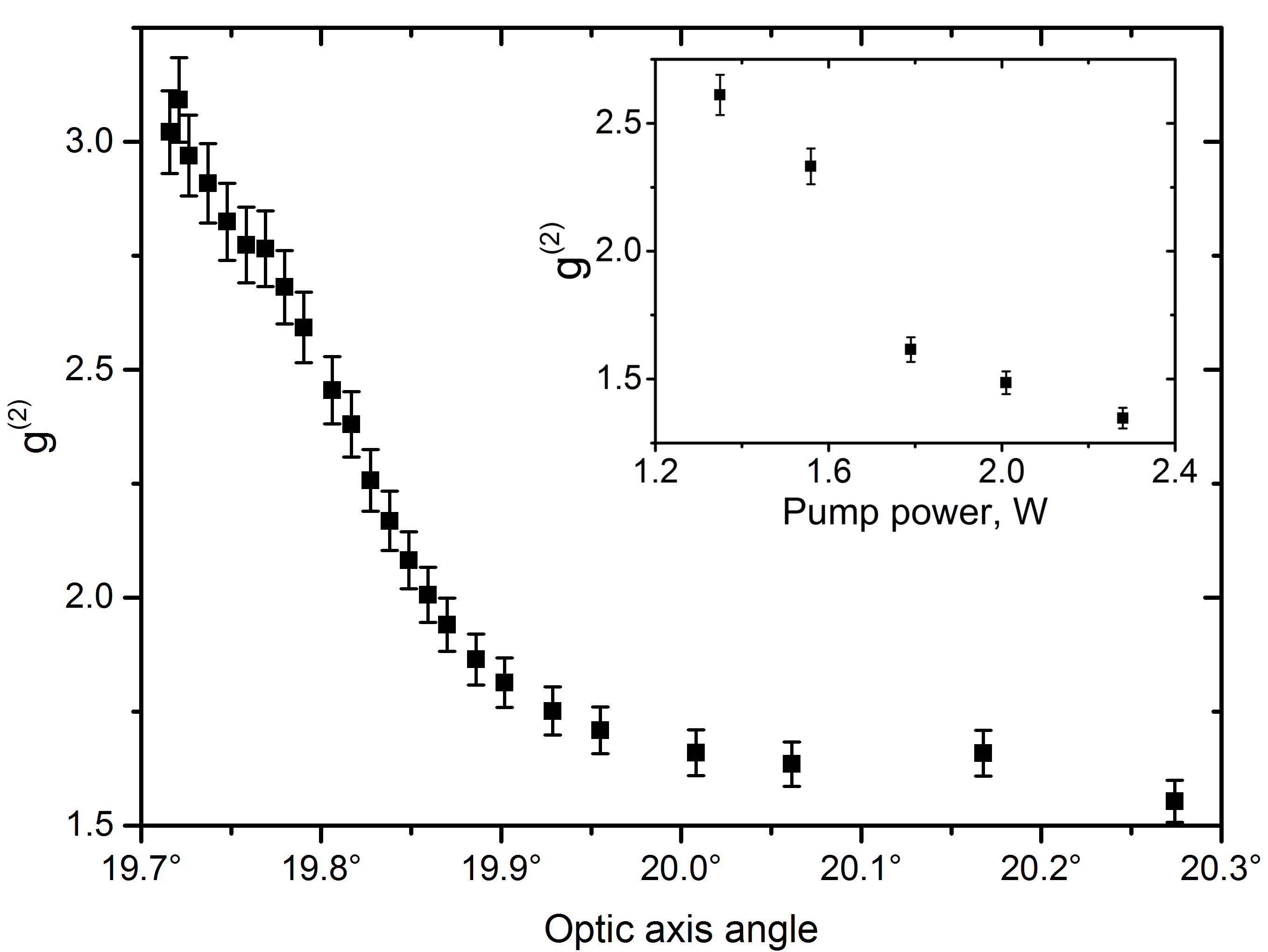}
\caption{The measured second-order correlation function $g^{(2)}$ for BSV depends on the angle of the BBO optic axis with respect to the pump direction, due to the unwanted second harmonic generated in the BBO crystal. Inset: due to the same effect, the $g^{(2)}$ also depends on the pump power.}
\end{center}
\end{figure*}

\begin{figure*}[h]
\begin{center}
\includegraphics[width=0.6\textwidth]{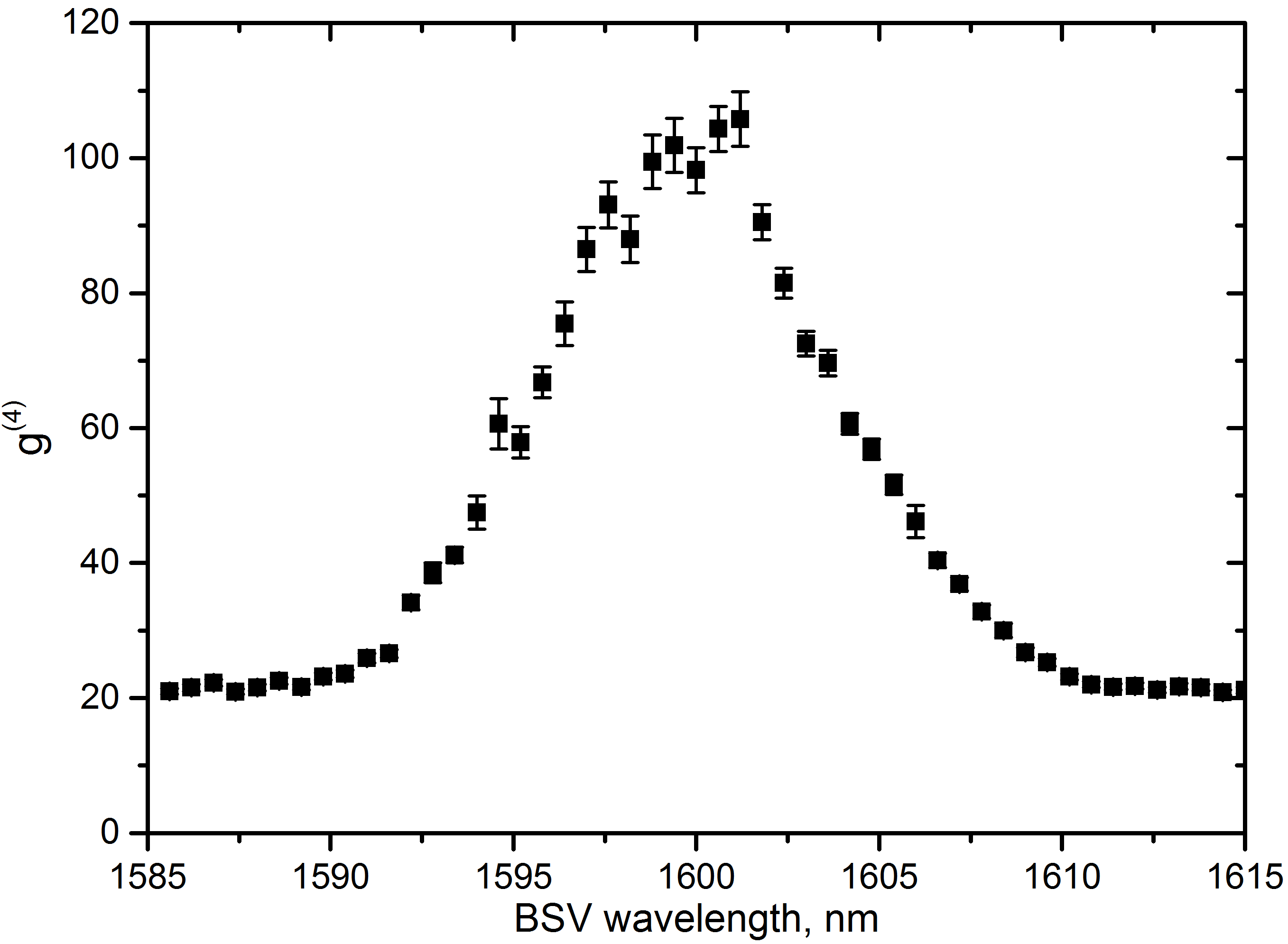}
\caption{The fourth-order correlation function $g^{(4)}$ versus the BSV wavelength.}
\end{center}
\end{figure*}

\begin{figure*}[h]
\begin{center}
\includegraphics[width=0.6\textwidth]{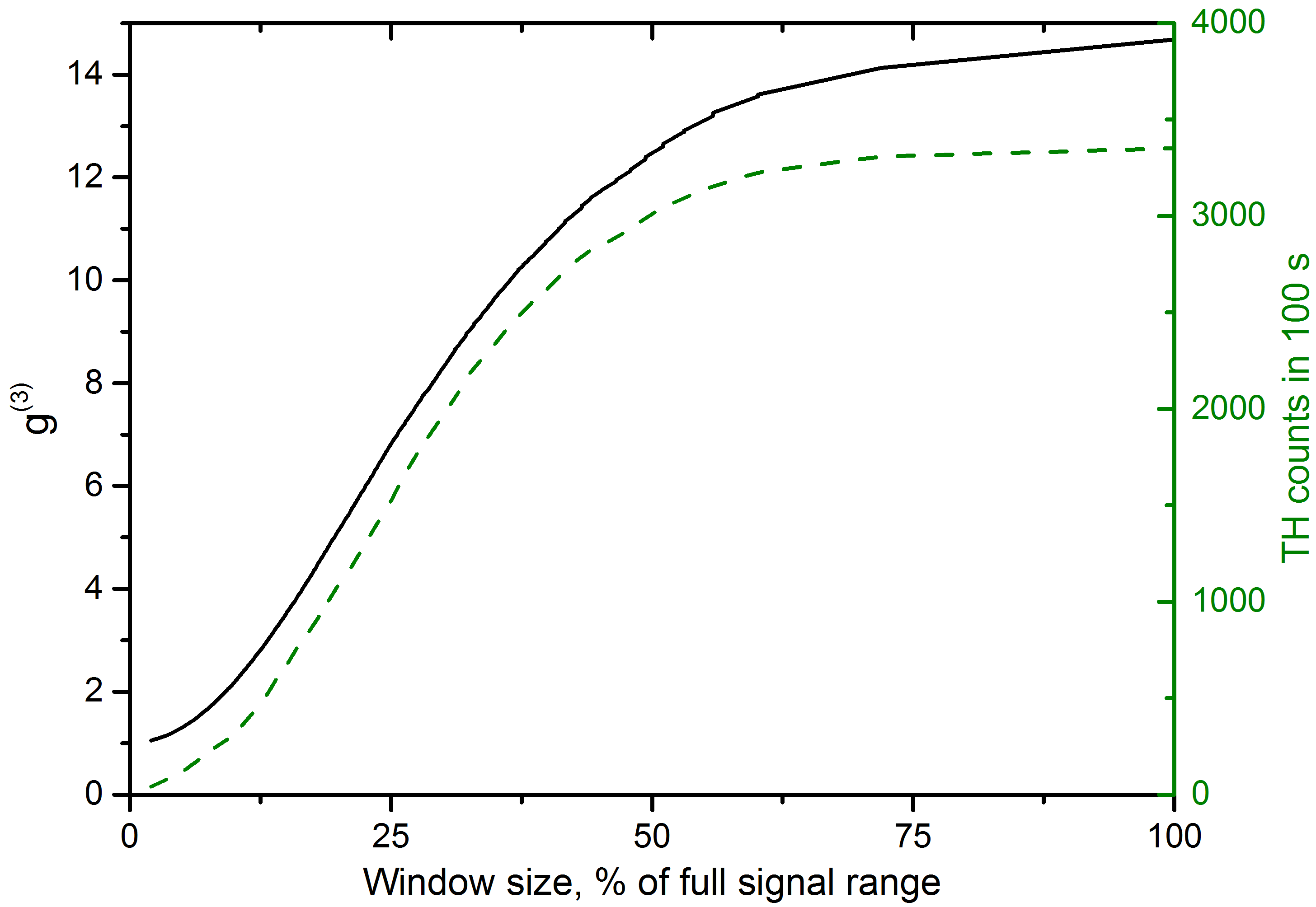}
\caption{The third-order correlation function $g^{(3)}$ for the post-selected data decreases almost to unity when the post-selection window for BSV signal becomes narrower (solid line). At the same time, the number of TH counts for a chosen pulses, within this window, decreases (dashed line). The low number of counts makes the harmonic flux measurable only with big errors.}
\end{center}
\end{figure*}

\begin{figure*}[h]
\begin{center}
\includegraphics[width=0.6\textwidth]{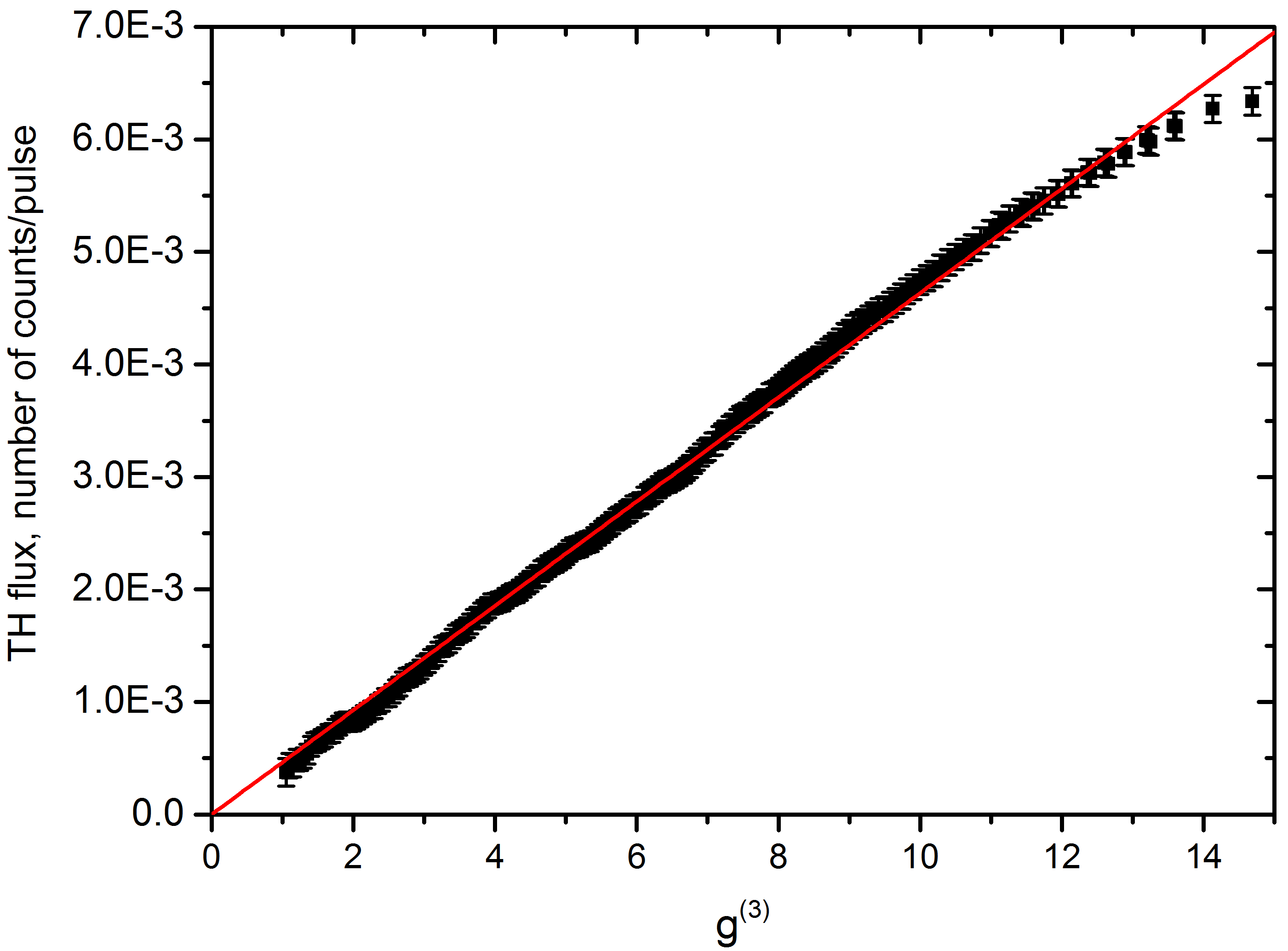}
\caption{The TH flux for post-selected pulses linearly depends on the third-order correlation function $g^{(3)}$.}
\end{center}
\end{figure*}


\begin{references}

\bibitem{Franken1961} Franken, P.A., Hill, A.E., Peters, C.W. \& Weinreich, G. Generation of Optical Harmonics. \textit{Phys. Rev. Lett.} \textbf{7} 118--119 (1961).

\bibitem{Denk1990}Denk, W., Strickler, J.H. \& Webb, W.W. Two-Photon Laser Scanning Fluorescence Microscopy. \textit{Science} \textbf{248}, 73--76 (1990).

\bibitem{Delone2000} Delone, N.B. \& Krainov, V.P. \textit{Multiphoton Processes in Atoms} (Springer, Berlin, 2000).

\bibitem{Gattass2008} Gattass, R.R. \& Mazur, E. Femtosecond laser micromachining in transparent materials. \textit{Nat. Photonics} \textbf{2} 219--225 (2008).

\bibitem{Shen1984} Shen, Y.R. \textit{The Principles of Nonlinear Optics}, (John Wiley \& Sons, New York, 1984).

\bibitem{Mollow1968} Mollow, B.R. Two-Photon Absorption and Field Correlation Functions. \textit{Phys. Rev.} \textbf{175} 1555--1563 (1968).

\bibitem{Agarwal1970} Agarwal, G.S. Field-Correlation Effects in Multiphoton Absorption Processes. \textit{Phys. Rev. A} \textbf{1} 1445--1459 (1970).

\bibitem{Lecompte1975} Lecompte, C., Mainfray, G., Manus, C. \& Sanchez, F. Laser temporal-coherence effects on multiphoton ionization processes \textit{Phys. Rev. A} \textbf{11} 1009--1015 (1975).

\bibitem{Delone1980} Delone, N.B. \& Masalov, A.V. Multiphoton detectors of laser radiation. \textit{Opt. Quant. Electron.} \textbf{12} 291--301 (1980).

\bibitem{Qu1992} Qu, Y. \& Singh, S. Photon correlation effects in second harmonic generation. \textit{Opt. Comm.} \textbf{90} 111-114 (1992).


\bibitem{Qu1995} Qu, Y. \& Singh, S. Measurements of photon statistics in second-harmonic generation. \textit{Phys. Rev. A} \textbf{51}, 2530--2536 (1995).

\bibitem{Popov2002} Popov, A.M. \& Tikhonova, O.V. The Ionization of Atoms in an Intense Nonclassical Electromagnetic Field. \textit{J. Exp. Theor. Phys.} \textbf{95}, 844--850 (2002).

\bibitem{Jechow2013}Jechow, A. \textit{et al.} Enhanced two-photon excited fluorescence from imaging agents using true thermal light. \textit{Nat. Photonics} \textbf{7}, 973--976 (2013).


\bibitem{Corzo2011}Corzo, N., Marino, A. M., Jones, K. M. \& Lett, P. D. Multi-spatial-mode single-beam quadrature squeezed states of light from four-wave mixing in hot rubidium vapor. \textit{Opt. Express} \textbf{19}, 21358--21369 (2011).

\bibitem{Jedrkiewicz2004}Jedrkiewicz, O. \textit{et al.} Detection of Sub-Shot-Noise Spatial Correlation in High-Gain Parametric Down Conversion. \textit{Phys. Rev. Lett.} \textbf{93}, 243601 (2004).


\bibitem{Iskhakov2009}Iskhakov, T.Sh., Chekhova, M.V. \& Leuchs, G. Generation and direct detection of broadband mesoscopic polarization-squeezed vacuum. \textit{Phys. Rev. Lett.}~\textbf{102}, 183602 (2009).

\bibitem{Agafonov2010}Agafonov, I.N., Chekhova, M.V. \& Leuchs, G. Two-color bright squeezed vacuum. \textit{Phys. Rev. A} \textbf{82}, 011801 (2010).

\bibitem{Glauber1963} Glauber, R.J. The Quantum Theory of Optical Coherence. \textit{Phys. Rev.} \textbf{130}, 2529--2539 (1963).


\bibitem{Perez2015}P\'{e}rez, A.M. \textit{et al.} Giant narrowband twin-beam generation along the pump-energy propagation direction. \textit{Nat. Commun.} \textbf{6}, 7707 (2015).

\bibitem{Spasibko2016}Spasibko, K.Yu. \textit{et al.} Ring-shaped spectra of parametric downconversion and entangled photons that never meet. \textit{Opt. Lett.} \textbf{41}, 2827--2830 (2016).

\bibitem{Janszky1987}Janszky, J. \& Yushin, Y. Many-photon processes with the participation of squeezed light. \textit{Phys. Rev. A} \textbf{36}, 1288--1292 (1987).


\bibitem{Boitier2011}Boitier, F. \textit{et al.} Photon extrabunching in ultrabright twin beams measured by two-photon counting in a semiconductor. \textit{Nat. Commun.} \textbf{2}, 425 (2011).

\bibitem{Iskhakov2012}Iskhakov, T.Sh. \textit{et al.} Superbunched bright squeezed vacuum state. \textit{Opt. Lett.} \textbf{37}, 1919--1921 (2012).

\bibitem{Strekalov2005} Strekalov, D., Matsko, A.B., Savchenkov, A.A., \& Maleki, L. Quantum-correlation metrology with biphotons: where is the limit? \textit{J. Mod. Opt.} \textbf{52}, 2233--2243 (2005).

\bibitem{Tiihonen2006} Tiihonen, M. \textit{et al.}. Ultrabroad gain in an optical parametric generator with periodically poled KTiOPO$_4$, \textit{Appl. Phys. B} \textbf{85}, 73--77 (2006).




\bibitem{Katamadze2015} Katamadze, K.G., \textit{et al.} Broadband biphotons in a single spatial mode. \textit{Phys. Rev. A} \textbf{92}, 023812 (2015).


\bibitem{Krasinski1976} Krasinski, J, \& Dinev, S. Influence of nonlinear effects on the statistical properties of a high power density laser beam. \textit{Optics Comm.} \textbf{18}, 424--426 (1976).

\bibitem{Leuchs1986} Leuchs, G. Photon Statistics, Antibunching and Squeezed States. \textit{NATO Science Series: B} \textbf{135}, 329--360 (1986).

\bibitem{Iskhakov2016} Iskhakov, T.Sh., \textit{et al.} Low-noise macroscopic twin beams, \textit{Phys. Rev. A} \textbf{93}, 043849 (2016).

\bibitem{Ducuing1964} Ducuing, J. \& Bloembergen, N. Statistical Fluctuations in Nonlinear Optical Processes. \textit{Phys. Rev.} \textbf{133}, A1493--A1502 (1964).

\bibitem{Teich1966} Teich, M.C. \& Wolga, G.J. Multiple-Photon Processes and Higher Order Correlation Functions. \textit{Phys. Rev. Lett.} \textbf{16}, 625--628 (1966).

\bibitem{Allevi2015} Allevi, A. \& Bondani, M. Direct detection of super-thermal photon-number statistics in second-harmonic generation. \textit{Opt. Lett.} \textbf{40}, 3089--3092 (2015).

\bibitem{Dudley2014} Dudley, J.M., Dias, F., Erkintalo, M., \& Genty, G. Instabilities, breathers and rogue waves in optics. \textit{Nat. Photonics} \textbf{8}, 755--764 (2014).

\bibitem{Meuret2015} Meuret, S., \textit{et al.} Photon Bunching in Cathodoluminescence. \textit{Phys. Rev. Lett.} \textbf{114}, 197401 (2015).

\bibitem{Rostovtseva1980} Rostovtseva, V.V., Saltiel, S.M., Sukhorukov, A.P., \& Tunkin, V.G.. Generation of higher optical harmonics in focused beams. \textit{Sov. J. Quantum Electron.} \textbf{10} 616--620 (1980).







\clearpage
\end{references}
\end{document}